\documentclass[12pt]{article}
\usepackage{epsfig}
\usepackage{aaspp4}

\begin{document}

\title{Probing Hierarchical Clustering by Scale-Scale 
    Correlations of Wavelet Coefficients}

\author{Jes\'us Pando\altaffilmark{1,2}
,
   Peter Lipa\altaffilmark{3}
,
Martin Greiner\altaffilmark{4} and 
Li-Zhi Fang\altaffilmark{1}
}

\altaffiltext{1}{Department of Physics, University of Arizona, Tucson,
AZ 85721}
\altaffiltext{2}{UMR 7550 CNRS, Observatoire de Strasbourg, 67000 Strasbourg
 France}
\altaffiltext{3}{Institut f\"ur Theoretische Physik, Technische
Universit\"at, D-01062 Dresden, Germany}
\altaffiltext{4}{Institut f\"ur Hochenergiephysik der \"Osterreichischen 
Akademie
der Wissenschaften, Nikolsdorfer Gasse 18, A-1050 Vienna, Austria} 

\begin{abstract}
It is of fundamental importance to determine if and how hierarchical
clustering is involved in large-scale structure formation of the
universe. Hierarchical evolution is characterized by rules which
specify how dark matter halos are formed by the merging of halos at smaller
scales.  We show that scale-scale correlations of the matter density
field are direct and sensitive measures to quantify this merging
tree.  Such correlations are most conveniently determined from
discrete wavelet transforms. Analyzing two samples of Ly$\alpha$
forests of QSO's absorption spectra, we find significant scale-scale
correlations whose scale dependence is typical for branching
processes. Therefore, models which predict a ``history'' independent
evolution are ruled out and the halos hosting the Ly$\alpha$ clouds
must have gone through a ``history'' dependent merging process during
their formation.
\end{abstract}


\section{Introduction}

The recent discoveries of the excess of faint blue galactic counts
(Lilly et al.\ 1995, Ellis et al.\ 1996) and a substantial population of
star forming galaxies at redshift $z\simeq$ 3--3.5 (Steidel et
al.\ 1996) may be taken as evidence of hierarchical structure
formation (Kauffmann \& White 1993; Lacey \& Cole 1993; Navarro, Frenk
\& White 1996). However, the predicted abundance of galaxies at higher
redshifts depends on, at the very least, models of  stellar
population synthesis and the IMF. Besides these abundances,
a more direct and independent detection of hierarchical evolution
is necessary to form a convincing argument for this scenario as well
as to discriminate among different models of structure formation.

Scenarios of hierarchical matter clustering are defined in terms of 
rules which determine how dark halos evolve from scale to scale. 
Measuring correlations of structures at different scales is a
direct and sensitive way to test the hypothesis of hierarchical
clustering.  We propose to detect this hierarchical structure by studying
correlations between coefficients of a wavelet decomposition of the
density field $\rho(x)$. We suggest relevant measures
to quantify such scale-scale correlations and to discriminate between 
different scenarios. Scale-scale correlation measures have
been shown to effectively reveal hierarchical characteristics of
energy transfer in turbulence and multiparticle physics 
(Yamada \& Ohkitani, 1991, Greiner, Lipa, \& Carruthers, 1995.)

The hierarchical clustering model of galaxy formation generally refers to 
the assumption that the irreducible correlation functions are described 
by the hierarchical relations $\xi_n = Q_n \xi_2^{n-1}$, where $\xi_n$ 
is the $n$-th order correlation function, and $Q_n$ are constants 
(White 1979). Obviously, if the hierarchical relations hold exactly, the 
two-point correlation function plus all $Q_n$ completely characterize 
the statistical features of galaxy formation, including higher order 
correlations such as scale-scale correlations. However, the hierarchical 
relations are only approximately fulfilled. Studies in turbulence in which 
fractal hierarchical relations only approximately hold showed that scale-scale 
correlations were still useful in describing statistical features 
(Yamada \& Ohkitani, 1991.) 
Therefore, scale-scale correlations will be useful in revealing 
deviations from the linked pair ansatz, and to discriminate among models 
including those satisfying the hierarchical relations approximately.

\section{Scale-scale correlations}
In order to assess the discriminative power of scale-scale correlations
we discuss two exactly solvable  models of hierarchical
clustering. The first is a Gaussian version of the block model
(Cole \& Kaiser 1988) employing an additive merging
rule, while the second is similar in spirit, but uses a multiplicative
merging rule (adapted from Meneveau \&
Sreenivasan 1987).  Both models give {\em identical\/} first and second
order statistics and are constructed in such a way as to reproduce the
experimental power spectrum. However, their different merging
rules imply a quite different structure of correlations beyond second
order. 

{\bf The Gaussian block model}. 
A block of mass $M_0$ distributed uniformly over a length $L$ is 
successively divided by a
factor two, giving $2^j$ blocks at mass scale $M_0/2^j$ at the
$j$-th iteration, while the length scale is $L/2^j$.  Each block is
labeled by the pair $(j,l)$, where $j=0,\ldots,J$ denotes the scale
(with some appropriate cut-off scale $J$) and $l=0,\ldots,2^j-1$ gives
the position of a block at scale $j$.  The mean mass density of all
blocks is $\overline{\rho}= M_0/L$.  The density
contrast is defined by $\epsilon(x)\equiv (\rho(x)-
\overline{\rho})/\overline {\rho}$.

A realization of the density contrast field is then generated
as follows: For the largest block $(0,0)$, the density contrast is
$\epsilon_{0,0} = 0 + \tilde{\epsilon}$, where $\tilde{\epsilon}$ is
drawn from a Gaussian distribution with variance $\Sigma$. For the two
blocks (1,0) and (1,1) at the next finer scale $j=1$, the two density
contrasts are respectively enhanced and diminished by an additive
factor $\tilde{\epsilon}_{0,0}$ which is again drawn randomly from a
Gaussian, but with a different variance $\Sigma_0$. Iterating down to
scale $J$, we obtain the additive merging rule of the Gaussian block
model:
\begin{eqnarray}
\label{gbm_merge_rule}
\epsilon_{j+1,2l} & = & \epsilon_{j,l}+ \tilde{\epsilon}_{j,l} \nonumber\\ 
\epsilon_{j+1,2l+1} & = & \epsilon_{j,l} -
\tilde{\epsilon}_{j,l}
\end{eqnarray}
where each $\tilde{\epsilon}_{j,l}$ is drawn independently from a
Gaussian with variance $\Sigma_{j}$ which may depend on the scale
$j$ but not on the position $l$. It is clear that in this scenario
$\tilde{\epsilon}_{j,l}$, describing the difference between density
contrasts on scales $j+1$ and $j$ at overlapping positions, does not
depend on the density $\epsilon_{j,l}$ in the parent block and is in
this sense ``history'' independent.

The finest scale distribution $\epsilon_{J,l}$ 
yields a realization of the density
distribution $\rho(x) = \overline{\rho}[1+ \epsilon(x)]$, where
$\epsilon(x)= \epsilon_{J,l}$ when $Ll/2^J \leq x < L(l+1)/2^J$.  This
$\rho(x)$ is then to be compared to the observed density
distribution on spatial resolution $\sim L/2^J$.  In order to
reproduce the power spectrum $P(k)$ one has to choose for the
variances $\Sigma^2$ and $\Sigma_j^2$
\begin{eqnarray}
\Sigma^2 & = &\frac{L}{2\pi} \int_{0}^{k_{0}} P(k) dk \\ \Sigma_j^2 &
= & \frac{L}{2\pi} \int_{k_j}^{k_{j+1}} P(k) dk, \ \ j=0,\ldots,J-1
\end{eqnarray}
where $k_j = \pi 2^{j+1} /L$.

Since all $\epsilon_{J,l}$ are constructed from sums of
independent Gaussian random variables, the resulting density matter
field $\rho(x)$ is a Gaussian random field which, by definition, 
produces no genuine correlations beyond second order. 

{\bf The branching block model}. 
To apply block models to dark halos, we should identify holes with 
blocks satisfying the condition of collapse. This procedure
introduces a history dependence in the sense that now the density
differences between adjacent scales do depend on the value of the
parent block.  Since we are not concerned with numerical details at
the moment, we simply simulate this history dependence as follows: For
block $(0,0)$, the density contrast is assigned just as in the
Gaussian block model.
 The mass $M$ of block
$(0,0)$ is split into blocks $(1,0)$ and $(1,1)$ with unequal
masses of $M(1+\alpha_0)/2$ and $M(1 -\alpha_0)/2$ or, with
equal probability, vice versa.  
Obviously, mass is conserved at each
evolution step. The density
contrasts are thus enhanced or diminished by multiplicative random weights
$1\pm\alpha_0$, e.g.\ $\epsilon_{1,0}=(1 + \alpha_0)
(1+\epsilon_{0,0})-1$ and $\epsilon_{1,1} = (1 - \alpha_0)
(1+\epsilon_{0,0})-1$.  Here only the sign of $\alpha_j$ is a random
variable, assigning the positive sign with probability $1/2$ to the
left and right sub-blocks respectively; the value of $\alpha_j$ is a
scale-dependent constant in the range $0 \leq \alpha_j\leq 1$.

The general recurrence relation between the density contrast on 
adjacent scales $j$ and $j+1$ is now
\begin{eqnarray}
\label{split_merge_rule}
(1+\epsilon_{j+1,2l})   & = & (1 \pm \alpha_{j})(1+\epsilon_{j,l}) \nonumber\\
(1+\epsilon_{j+1,2l+1}) & = & (1 \mp \alpha_{j})(1+\epsilon_{j,l}) ,
\end{eqnarray}
describing a hierarchical branching process.
The multiplicative structure of this merging rule
renders the resulting density field $\rho(x)$
non-Gaussian and leads to significant genuine correlations beyond
second order.  The difference between density contrasts on scales $j$
and $j+1$, given by $\tilde{\epsilon}_{j, l} = \pm
\alpha_{j}(1+\epsilon_{j,l})$, is now obviously history dependent.

Although the Gaussian- and branching-block models have very different
evolution rules (\ref{gbm_merge_rule}) and
(\ref{split_merge_rule}), their first and second order moments can be made
equal if $\alpha_j$ is chosen recursively by
\begin{equation}
 \alpha_j^2 = \frac{\Sigma_j^2}{(\overline\rho^2+\Sigma^2) 
	(1 + \alpha_{0}^2) (1 + \alpha_{1}^2)\ldots (1 + \alpha_{j-1}^2)}.  
\end{equation}

We thus arrive at two paradigmatic models with quite different merging
scenarios with identical means and covariances. No measure
based on first and second order statistics can discriminate between
these scenarios.  Clearly, what is needed is a measure of the
hierarchical evolution by scale-scale correlations involving moments
of order higher than two.

The essential information of cluster formation  
is captured in the properties of the differences $\tilde{\epsilon}_{j,l}$
at two adjacent evolution steps. This is exactly the information
obtained by a discrete wavelet transform (DWT). More
generally, the coefficients $\tilde{\epsilon}_{j,l}$ of
{\em any\/} decomposition  
\begin{equation}
\epsilon(x) = \sum_{j=0}^{J-1} \sum_{l=0}^{2^j-1}
\tilde{\epsilon}_{j,l}\; \psi_{j,l}(x)
\end{equation}
with respect to a complete and orthogonal wavelet basis $\psi_{j,l}(x)$
(Daubechies 1992) provide similar
information on density differences at adjacent scales; 
for the purpose of analysis, the particular choice of wavelet is of
secondary importance and all will lead to comparable results 
(Pando \& Fang 1996).

The counts-in-cell (CIC) method has been applied to study the scale-dependence
of clustering and has even been applied to hierarchical studies (Balian \& 
Schaeffer 1989, Bromley 1994). However, there are differences 
between the CIC and DWT analysis. The basis of the DWT decomposition 
[eq.(6)] $\psi_{j,l}(x)$ is orthogonal with respect to both $j$ (scale) and 
$l$ (position), while the basis (or window) for the CIC is orthogonal with 
respect to $l$, but not to $j$. Moreover, the Haar wavelet scaling functions 
(which are equivalent to the CIC with cubic cells) are not localized in 
Fourier (scale) space. It is not possible to effectively measure scale-scale 
correlations by a scale-mixed decomposition.  On the other hand, 
hierarchical clustering is characterized by local relations between large and 
small structures, and therefore, the decomposition should also be localized in
physical space. The DWT is constructed by an orthogonal, complete and 
localized (in both physical and Fourier space) basis.  For these reasons, 
studying scale-scale correlations via the DWT is an effective tool.

The statistical properties of the wavelet coefficients
$\tilde{\epsilon}_{j,l}$ of a random field $\epsilon(x)$ are most
conveniently obtained once the generating function
\begin{equation}
Z^{(J)}[{\mathbf\eta}] =\left \langle \exp \left(
    i \sum_{j=0}^{J-1} \sum_{l=0}^{2^j-1} \eta_{j,l}\;\tilde\epsilon_{j,l}
       \right) \right \rangle
\end{equation}
is known,
where $\mathbf\eta$ represents the auxiliary variables $\eta_{j,l}$ and
$\langle \ldots \rangle$ denotes the ensemble average.
For both block model versions discussed above one can directly translate
the merging rules into recursion relations for their generating functions
$Z^{(j+1)}$ and $Z^{(j)}$ at two adjacent scales. The explicit formulae
may be found in more detailed expositions (Greiner, Lipa, \&
Carruthers 1995; Greiner et al.\ 1996)

Thus, various correlation quantities can be calculated from 
$Z^{(J)}[{\mathbf\eta}]$ by
taking appropriate derivatives. For instance, the correlations between
a wavelet coefficient of a block $l$ at scale $j$ and it's left sub-block $2l$
at scale $j+1$ are found by
\begin{equation}
\langle \tilde\epsilon_{j,l}^p \, \tilde\epsilon_{j+1,2l}^q \rangle
 =\left . \frac{1}{i^{p+q}}
\frac{\partial^{p+q}Z^{(J)}}{\partial\eta_{j,l}^p \, 
      \partial\eta_{j+1,2l}^q} \right |_{{\mathbf\eta} = 0} \ .
\end{equation}
Specifically, we use symmetric and normalized correlation measures with
even orders $p=q$,
henceforth called scale-scale correlations:
\begin{equation}
\label{sscorr}
C_j^{p,p} = 
\frac {2^{j+1}\sum_{l=0}^{2^j-1}\langle \tilde\epsilon_{j,l}^p \,
 \tilde\epsilon_{j+1,2l}^p \rangle}
{\sum_{l=0}^{2^j-1}\langle \tilde\epsilon_{j,l}^p \rangle \,
 \sum_{l'=0}^{2^{j+1}-1}\langle \tilde\epsilon_{{j+1},l'}^p \rangle } \ .
\end{equation}

As expected, for the Gaussian block model we obtain
$C_{j}^{p,p} = 1$ for all $p\geq 2$, i.e.\ there are
no scale-scale correlations of order greater than two. 
Higher order correlations for the branching block model have been 
calculated in Greiner et al.\ (1996); the $p=2$ scale-scale correlations
for a simplified version with scale-independent $\alpha_j = \alpha$ are
\begin{equation}
C_{j}^{2,2}= \frac{(1+6\alpha^2+\alpha^4)^j}{(1+\alpha^2)^{2j}} \ .
\end{equation}
Thus, the $C_{j}^{p,p}$ provide a sensitive quantification of
correlations between structures living at adjacent scales and
discriminate clearly between a Gaussian and branching scenario.

It is interesting to point out that if the hierarchical relations 
(White 1979) hold, the constant $Q_n$ can approximately be 
described by the scale-scale correlations. For instance, for n=4, we have
\begin{equation}
C^{2,2}_j \simeq Q_4 \left (\frac{1}{2^j} \sum_{l=0}^{2^{j}-1}
\langle \tilde\epsilon_{j,l}^2 \rangle 
+ \frac{1}{2^{j+1}}\sum_{l=0}^{2^{j+1}-1}
\langle \tilde\epsilon_{j+1,l}^2 \rangle \right ) .
\end{equation}
This relation shows that it is possible to test the assumption that the $Q_n$ 
are scale (or $j$)-independent by  scale-scale correlations 
(Pando, J. et al. 1997.)

\section{An example: Ly$\alpha$ absorption forests}

It is generally believed that the Ly$\alpha$ forests of QSO absorption spectra 
are due to the absorption of pre-collapsed clouds in the density field of 
the universe (Fang et al.\ 1996). Hierarchical clustering  requires that both 
collapsed halos and pre-collapsed clouds undergo similar merging evolutions. 
As such, the Ly$\alpha$ forests should be good candidates for detecting 
hierarchical clustering. 

We looked at two data sets of Ly$\alpha$ forests. The first was
compiled by Lu, Wolfe and Turnshek (1991, hereafter LWT). The total
sample contains $\sim$ 950 lines from the spectra of 38 QSO that
exhibit neither broad absorption lines nor metal line systems. The
second set is from Bechtold (1994, hereafter JB), which contains a
total of $\sim$ 2800 lines from 78 QSO spectra, in which 34 high
redshift QSO's were observed at moderate resolution. To eliminate the
proximity effect, all lines with $z \geq z_{em} - 0.15$ were deleted
from our samples (Pando \& Fang 1996). 
These samples cover a redshift range of 1.7 to 4.1,
and a comoving distance range from about $D_{\scriptstyle min}$=
2,300 $ h^{-1}$Mpc to $D_{\scriptstyle max}=$3,300 
$h^{-1}$Mpc, if $q_{0} = 1/2$ and h$=H_0/100$ km s$^{-1}$ Mpc$^{-1}$.

We make block trees from the largest block $L=D_{\scriptstyle max}-
D_{\scriptstyle min}$ with $L/2^j$, 
and $j=0,\ldots,9$. The smallest block-size $L/2^9 \sim 2$ h$^{-1}$ 
Mpc is about the scale where the effect of line blending occurs.
Moreover, since we will only study scale-scale correlations on scales
equal to or larger than about $L/2^8 \sim 5$ h$^{-1}$ Mpc, the influence of
peculiar motions should be negligible.

To reduce the influence of the $z$-dependence in the mean density 
$\overline\rho$  of Ly$\alpha$ lines, we chop the entire red-shift space
into segments with size $\bigtriangleup z = 0.4$. This corresponds to a
comoving space of 270 h$^{-1}$Mpc for the lowest $z$ of the LWT and JB
samples, and 110 h$^{-1}$Mpc for the highest $z$ of the samples.

To account for the remaining $z$-dependence of  
$\overline\rho$,  100 random samples
for each data set are generated by shifting each observed line by a random 
distance $\delta D$ not exceeding the interval distance corresponding 
to $\bigtriangleup z$. Any line shifted outside the interval is wrapped around 
to bring it back into the interval. This procedure gives a de-correlated
(random) sample which still reflects the $z$-dependence of the observed
sample. Both, the observed and random samples are 
suitable for statistical analysis on scales less than 100 h$^{-1}$ Mpc.

The real data and random sample are subjected to the four-coefficient 
Daubechies discrete wavelet (D4), which is better localized in Fourier space
than the Haar wavelet.
Fig.\ 1 shows the results for $C_{j}^{2,2}$ of the
LWT and JB samples with line widths $> 0.32$ \AA. 
Clearly, the scale-scale correlation
$C_{j}^{2,2}$ for the observational data is significantly larger than unity and
well above the random samples on all scales $j \geq 5$ (i.e. less than 
about 80 h$^{-1}$ Mpc). 
More importantly,  the two independent data sets, LWT and JB, show 
similar behavior. Thus, the detected 
scale-scale correlations seem to be an intrinsic feature of the 
clustered density field traced by Ly$\alpha$ forests. 
The influence of the $z$-dependence of $\overline\rho$ is estimated by 
the values of $C_{j}^{2,2}$ for the random samples (shaded regions in
Fig.\ 1); these are slightly above unity, but can certainly not explain the
strong $j$-dependence of the LWT and JB samples.

The Ly$\alpha$ absorption line distribution is a discrete process. The
discreteness is a source of non-Gaussianity as it is very well known
that Poisson noise is non-Gaussian.  The question naturally arises as to
whether the non-Gaussianity measured by
$C_{j}^{p,q}$ is caused by a Poisson process. This non-Gaussianity
has been carefully studied in Greiner, Lipa and Carruthers (1995) and Fang
and Pando (1997). The main conclusion is that the non-Gaussianity of
Poisson noise is significant only on scales of the mean distance of nearest 
neighbors. The distributions on large scales are a superposition of 
the small scale field. According to the central limit theorem 
the non-Gaussianity of Poisson noise will rapidly and monotonously approach 
zero on larger scales. In our analysis the scales being studied are larger 
than 5 h$^{-1}$ Mpc which is much larger than the mean distance between
nearest neighbor Ly-alpha lines. $C^{p,q}_j \neq 1$ is not
due to the discreteness of samples or noise, especially for
larger scales.

Fig.\ 1 also demonstrates that the 
branching block model reproduces the trend
of the observed data, while the Gaussian block model ($C_{j}^{2,2}\equiv 1$)
certainly lacks a mechanism that produces the observed hierarchical correlation
structure even when the $z$-dependence of the
mean number density of Ly$\alpha$ clouds is taken into account. 

\begin{figure}[ht]
\begin{center}
\epsfig{file=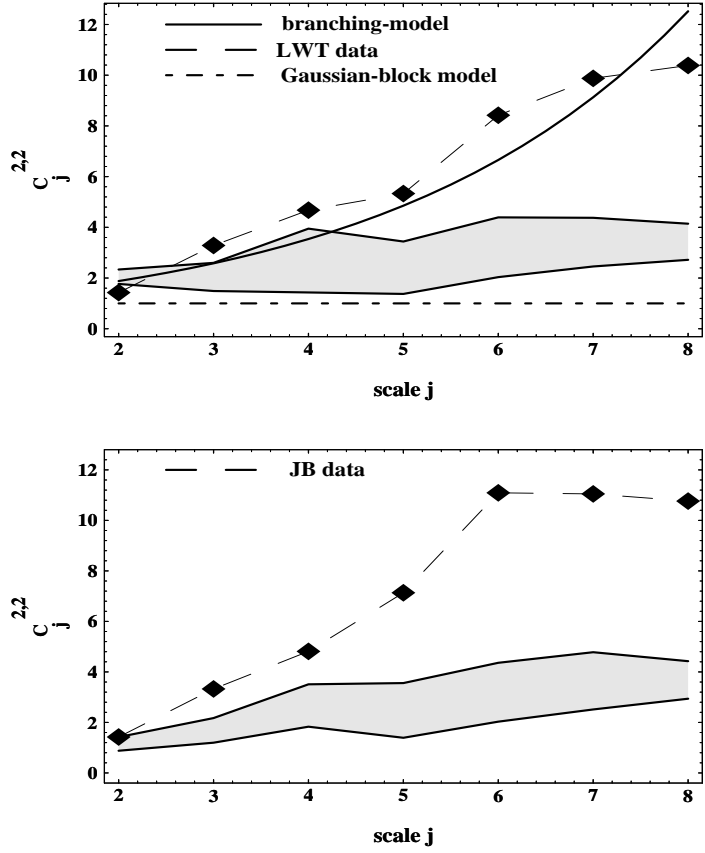,height=5in,width=4in}
\label{fig1}
\caption{
Scale-scale correlations $C_{j}^{2,2}$  for Ly$\alpha$ samples 
LWT (top) and JB (bottom) with line width larger than $0.32$\AA.
The curves show the values for the Gaussian block model 
(dash-dotted) and the branching block model (full line) with $\alpha=0.34$.
The shaded areas are the $\pm\sigma$ of 100 random samples for
the LWT and JB cases respectively. }
\end{center}
\end{figure}

\section{Conclusion}

Scale-scale correlations $C_{j}^{p,p}$ and possible variants are viable
statistical measures to discriminate between different scenarios of
merging dynamics of large scale structures of the universe. They
provide a direct clue of how larger halos and clouds are built up from their
substructures at smaller scales. The wavelet transform is a fast and 
convenient tool to obtain the necessary information on
localized contributions to the matter density field at different
scales.

Just as the two-point correlation function $1+\xi(x) > 1$ is an indicator of 
spatial clustering, scale-scale correlations $C_{j}^{p,p}>1$ 
indicate some sort of ``history'' dependence in hierarchical 
clustering schemes. 
The scale-scale correlations of the one-dimensional 
Ly$\alpha$ forests show, indeed, features expected by multiplicative
hierarchical clustering. Similar features have been found in 
the case of hydrodynamical turbulence.
One can conclude that the halos hosting Ly$\alpha$ clouds 
must have undergone a ``history'' dependent evolution process in 
some way during their formation. 

Finally, it is not difficult to generalize this method to three dimensions.

\acknowledgments

M.G.\ acknowledges support from GSI, DFG and BMBF. P.L.\ is
indebted to the \"Osterreichische Aka\-de\-mie der Wissenschaften for
support with an APART fellowship. The authors thank the referee for helpful 
comments, especially on the interesting question of the relation 
between $Q_n$ and $C^{p,p}_j$.

\clearpage

\end{document}